\documentclass[12pt]{article}

\usepackage{bm}
\usepackage{amssymb}
\usepackage{enumerate}

\usepackage{graphicx} 
\begin{document}

\title{Nonlocal   $SU(5)$ GUT}
\author{N.V.Krasnikov \thanks{ Nikolai.Krasnikov@cern.ch}
\\Institute for Nuclear Research RAS\\
\\ Moscow 117312, Russia
\\and
\\ JINR Dubna, Russia
}
\maketitle
\begin{abstract}

We show that in  nonlocal generalization of standard nonsupersymmetric 
 $SU(5)$ GUT  it is possible to 
solve the problems with the proton lifetime and the Weinberg angle 
without introduction of additional particles in the spectrum of the theory. 
Nonlocal scale $\Lambda$ responsible for ultraviolet cutoff 
coincides (up to some factor) with GUT scale $M_{GUT}$. 
We find that in the simplest nonlocal modification of the 
$SU(5)$ model  $M_{GUT} \approx 3 \cdot 10^{16}~GeV$. In general 
case the value of $M_{GUT}$ is an arbitrary and the most interesting option 
$M_{GUT} = O(M_{PL})$ could be realized.

\end{abstract}
\newpage

The remarkable success of the supersymmetric  $SU(5)$ grand unified theory (GUT) \cite{1}-\cite{17} 
was considered by many physicists as the first hint in favour of the existence 
of low energy broken supersymmetry in nature. 
However the nonobservation of supersymmetry at the LHC is probably the opposite 
hint that the supersymmetry concept and in particular the supersymmetric $SU(5)$ 
GUT is wrong. 
It is well known  that the 
standard $SU(5)$ GUT  \cite{18}  is in conflict with experimental data
\cite{13,14}.  
So a  natural question arises: is it possible to invent 
nonsupersymmetric generalizations of the standard $SU(5)$ GUT 
non contradicting to  the experimental data?  The answer is positive, in particular,
 in the $SO(10)$ GUT the introduction of 
the intermediate scale $M_I \sim 10^{11} GeV$ allows to obtain the Weinberg 
angle $\theta_w$ in agreement with experiment \cite{19}. In Refs.\cite{20,21} 
    the introduction of the additional split multiplets $5 \oplus \overline{5}$  
and $10 \oplus \overline{10}$ 
  in  the $SU(5)$ model has been proposed. 
In Ref.\cite{22} the extension of the standard $SU(5)$ GUT with 
light scalar colour octets and electroweak triplets has been considered. 

In this note we point out that  in nonlocal generalization of $SU(5)$ GUT  
it is possible to solve the problems with the proton lifetime and the Weinberg angle 
by the introduction of additional nonlocal  terms in the Lagrangian that  
leads   to the modification of the GUT condition 
$\alpha_1(M_{GUT}) =  \alpha_2(M_{GUT})  = \alpha_3(M_{GUT})$ for the effective coupling constants.  
Nonlocal scale $\Lambda$ responsible for ultraviolet cutoff 
coincides (up to some factor) with GUT scale $M_{GUT}$. 
In the simplest nonlocal  modification of the standard renormalizable 
$SU(5)$ GUT  the value of the GUT scale  is $M_{GUT} \approx 3 \cdot 10^{16}~GeV$.   
 In general 
case the value of $M_{GUT}$ is an arbitrary and the most interesting option 
$M_{GUT} = O(M_{PL})$ could be realized.

Let us start with the   observation that in 
standard $SU_c(3)\otimes SU_L(2) \otimes U(1)$ gauge  model the effective 
coupling constants $\alpha_3(\mu)$ and $\alpha_2(\mu)$ cross each other
( $\alpha_3(M_{GUT})  = \alpha_2(M_{GUT})$)
 at the 
scale $M_{GUT}  \approx O(10^{17}~GeV)$. At one-loop level 
the effective coupling constants $\alpha_i(\mu)$ obey the equations
\begin{equation}
\mu\frac{d\alpha_i(\mu)}{d\mu} = \frac{b_i}{2\pi}\alpha^2_i(\mu) \\,
\end{equation}
where for the SM model with 3 generations $b_3 = -7$, $b_2 = -3\frac{1}{6}$ 
and $b_1 = 4.1$. 
As a consequence  we find  that
\begin{equation}
\frac{1}{\alpha_2(m_t)} - \frac{1}{\alpha_3(m_t)} = 
\frac{b_2 -b_3}{2\pi}\ln(\frac{M_{GUT}}{m_t}) \\.
\end{equation}
Numerically   $M_{GUT} = (0.9 \pm 0.2) \cdot 10^{17}~GeV$ 
and $\frac{1}{\alpha_3(M_{GUT})} = 46.9 \pm 0.2$ 
\footnote{
In our estimates we use $\alpha_3(m_Z) = 0.118 \pm 0.001$, 
 $\sin^2(\theta_W)(m_Z) = 0.231 \pm 0.001 $ and $\alpha^{-1}_{em}(m_Z) = 127.8 \pm 0.1 $.
}.

The unification scale   $M_{GUT}  =  (0.9 \pm 0.2) \cdot 10^{17}~GeV$ 
  is safe  for    the current proton decay bound \cite{Bound}. 
Really, in  standard $SU(5)$ model the proton lifetime 
due to the massive vector exchange is determined by the formula \cite{Proton}
\begin{equation}
\Gamma(p \rightarrow e^{+} \pi^{o})^{-1} = 4 \cdot 10^{29 \pm 0.7}
(\frac{M_v}{2 \cdot10^{14} Gev})^{4} ~yr \,,
\end{equation}
where   $M_v \equiv  M_{GUT} = \sqrt{\frac{5}{24}}g_5 \Phi_0$ is the mass 
of vector bosons responsible for proton decay\footnote{Here $\Phi_0$ is 
the vacuum expectation value of the $SU(5)$ scalar 24-plet  
$<\Phi> = \frac{\Phi_0}{\sqrt{15}}Diag(1,1,1 -3/2. -3/2)$ 
responsible for $SU(5) \rightarrow  SU_c(3)\otimes SU_L(2)\otimes U(1)$ 
gauge symmetry breaking and $g_5$ is the $SU(5)$ gauge coupling at the GUT scale $M_{GUT}$.} . 
From the current experimental limit \cite{Bound}
$\Gamma(p \rightarrow e^{+} \pi^{o})^{-1} \geq 1.67 \cdot 10^{34} ~yr $ 
we conclude that $M_{GUT} \geq 2.5 \cdot 10^{15}~ GeV$. 
The main problem of the standard SU(5) GUT   
with the unification scale $M_{GUT} \approx   10^{17}~GeV$
is that 
the  experimental  values of $\alpha_3(m_Z), ~\sin^2(\theta_W)(m_Z), ~\alpha^{-1}_{em}(m_Z)$ 
lead to 
non equal values of the effective coupling constants  $\alpha_3(M_{GUT})$ and   
$\alpha_1(M_{GUT})$, namely 
$\alpha^{-1}_1(M_{GUT}) = 36.0 \neq \alpha^{-1}_3(M_{(GUT}) = 46.9 $. 

Our main observation is that 
the use of nonrenormalizable interaction\footnote{In Refs.\cite{X,Y} the influence of 
nonrenormalizable 
interaction  $L_{nl} = \frac{c}{M_{PL}}Tr(F_{\mu\nu}\Phi F^{\mu\nu})$ with $c = O(1)$  has been studied. 
It was realized that this interaction allows to increase the GUT scale but can't 
solve the problem with wrong Weinberg 
angle prediction.}
\begin{equation}
\Delta L_{F \Phi F \Phi} = \frac{1}{4\Lambda_{\Phi1}^2}(Tr(F_{\mu\nu}\Phi))(Tr(F^{\mu\nu}\Phi))  
\end{equation}
leads to the additional term for the effective coupling constant $\alpha_1(\mu)$ 
at GUT scale, namely 

\begin{equation}
 \frac{1}{\alpha_1(M_{GUT})} = 
\frac{1}{\alpha_{3}(M_{GUT}}) - \Delta \\ ,
\end{equation}
where
\begin{equation}
\Delta = \frac{\pi \Phi^2_o}{\Lambda^2_{\Phi1}} = 
\frac{1}{\alpha_3(M_{GUT})}\frac{6 M^2_v}{5\Lambda^2_{\Phi1}} \\ .
\end{equation}
Numerically we obtain  $\Delta = 10.9 \pm 0.2$ and $ \Lambda_{\Phi 1} \approx 2.3 \cdot M_v$.  
 So we find  that additional nonrenormalizable interaction (4)  modifies 
GUT unification condition  in such a way that 
 the unification takes place at the scale  $M_{GUT} \approx  10^{17}~GeV$ 
nondangerous for proton decay bound and   the unification scale $M_{GUT}$ 
does not contradict to  the  experimental values of  $\sin^2(\theta_W)(M_Z)$ and $\alpha^{-1}(M_Z)$.
The appearance of additional arbitrary parameter $\Delta $ in the relation (5) means 
that we can't predict the value of $\sin^2(\theta_W)$. Here  the  untrivial fact is that 
the unifcation of $\alpha_2(\mu)$ and   $\alpha_3(\mu)$ effective coupling constants 
takes place at the scale  $M_{GUT} = O(10^{17}~GeV)$ which is safe 
for the  proton lifetime  bound. An account of two-loop effects for the 
evolution of the effective couplings $\alpha_k(\mu)$  
  leads \cite{11} to the replacement 
\begin{equation}
\frac{1}{\alpha_k(m_Z)} \rightarrow \frac{1}{\alpha_k(m_Z)} - \theta_k  \\,
\end{equation}
where
\begin{equation}
\theta_{k} = \frac{1}{4\pi}\sum_{j=1}^{3} 
\frac{b_{kj}}{b_{j}}ln[\frac{\alpha_{j}(M_{GUT}}
{\alpha_{j}(m_Z)}]  \\.
\end{equation}
Here $b_{ij}$ are the two-loop $\beta$-functions  coefficients
\footnote{
At two loop level the renormalization group equations for $\alpha_i(\mu)$ effective 
coupling constants are $\mu\frac{d\alpha_i}{d\mu} = 
\frac{b_i}{2\pi}\alpha^2_i + \sum_{j=1}^{j=3}\frac{b_{ij}}{4\pi^2}\alpha^2_i\alpha_j$, 
see Ref.\cite{two1, two2}.  }. 
An account of two-loop corrections leads to the   decrease  of  
$M_{GUT}$ by factor 3. 
The parameter $\Delta$ in (5) is not small.
Really, $  \Delta/(\frac{1}{\alpha_2(M_{GUT})}) \approx  0.24    $  and 
$ \Lambda_{\Phi 1} \approx 2.3 \cdot M_v$.  
It means that at the scale $M_{GUT}$ we must have some ultraviolet cutoff(regulator) 
to make sence to  the nonrenormalizable interaction (4) at quantum level.
Probably the most 
promising  way to deal with nonrenormalizable theories is the use of 
nonlocal field theory \cite{EFIMOV1, EFIMOV2}. The simplest nonlocal generalization 
  of the   renormalizable Yang-Mills Lagrangian 
\begin{equation} 
L_{YM} = -\frac{1}{2 g^2_5}Tr(F_{\mu\nu}F^{\mu\nu}) \\,
\end{equation}
is \cite{KRASNIKOV} 
\begin{equation} 
L_{YM,nl} = -\frac{1}{2g^2_5}Tr(F_{\mu\nu}V(-\Delta_{\mu}\Delta^{\mu})   F^{\mu\nu}) \\,
\end{equation}
where $
F_{\mu\nu}  = \Delta_{\mu}A_{\nu} - \Delta_{\nu}A_{\mu} $,  $\Delta_{\mu} = \partial_{\mu} 
-i A_{\mu} $,   $A_{\mu} = A^a_{\mu}T_a$\footnote{Here   $T_a$ are the $SU(5)$ 
matrices  with  $Tr(T_aT_b) = \frac{1}{2}\delta^a_b$ and $g_5$ is the $SU(5)$ 
gauge coupling constant.} and 
the  formfactor $V(x)$ is entire function  on $x$. 
The  gauge propagator $D_{\mu\nu}^{nl}(p^2)$ for the nonlocal Lagrangian (10) 
in Feynman gauge 
is 
\begin{equation}
D^{nl}_{\mu\nu}(p^2) = \frac{g_{\mu\nu}}{i g^2_5 \cdot p^2} \cdot \frac{1}{V(p^2)} \,.
\end{equation}
The use of nonlocal formfactor $V(p^2)$ with decreasing behaviour 
in the euclidean region at  $p^2 \rightarrow -\infty$, 
for instance $V^{-1}(p^2) = \exp(p^2/\Lambda_{\Phi1})$
makes the Yang-Mills model superrenormalizable \cite{KRASNIKOV}\footnote{We can consider 
nonlocal Yang-Mills Lagrangian (10) as a generalization of 
Slavnov gauge invariant  regularization \cite{SLAVNOV1, SLAVNOV2} of Yang-Mills model  with higher 
order derivatives.}.
Possible nonlocal generalization of nonrenormalizable interaction (4) is 
\begin{equation}
\Delta L_{F \Phi F \Phi,nl} = -\frac{1}{4\Lambda_{\Phi1}^2}
(Tr(F_{\mu\nu}\Phi)   V_{\Phi1}(-\partial^{\mu}\partial_{\mu})    (Tr(F^{\mu\nu}\Phi))  
\end{equation}
with $ V_{\Phi1}(p^2) \sim \exp(p^2/\Lambda_{\Phi1})$
The use of nonlocal formfactors $V $ and $V_{\Phi 1}$ cures bad ultraviot 
properties of nonrenormalizable interaction (4) and make it superrenormalizable.
For nonlocal Lagragian (12)  the parameter $\Delta$ in formula (5) depends on  the 
scale $\mu$
\begin{equation}
\Delta(\mu) =  \frac{\pi \Phi^2_o}{\Lambda^2_{\Phi1}}V_{\Phi 1}(-\mu^2)
\end{equation}
We can use  the normalization condition $ V_{\Phi 1}(-M^2_{GUT}) =1$. In this case formula (6) 
and numerical estimate  for $\Delta$ are valid.


We can also add  to the $SU(5)$ Lagrangian  other nonrenormalizable term 
\begin{equation}
\Delta L_{F\Phi F \Phi2} = -\frac{Tr(F_{\mu\nu}\Phi^2F^{\mu\nu})}{4\Lambda^2_{\Phi2}} \,.
\end{equation}
Nonzero vacuum expectation value  $<\Phi> =  
\frac{\Phi_0}{\sqrt{15}}Diag(1,1,1 -3/2. -3/2 )$  of $SU(5)$ 24-plet $\Phi$
 leads to additional contributions 
for  coupling constants at GUT scale, namely 
\begin{equation}
\frac{1}{\alpha_1(M_{GUT})} \rightarrow \frac{1}{\alpha_1(M_{GUT})} + \frac{7\kappa}{120} \\,
\end{equation}
\begin{equation}
\frac{1}{\alpha_2(M_{GUT})} \rightarrow \frac{1}{\alpha_2(M_{GUT})} + \frac{3\kappa}{40} \\,
\end{equation}
\begin{equation}
\frac{1}{\alpha_3(M_{GUT})} \rightarrow \frac{1}{\alpha_3(M_{GUT})} + \frac{\kappa}{30} \\,
\end{equation}
where $\kappa = \frac{4\pi \Phi^2_0}{\Lambda^2_{\Phi2}}$. As a consequence we find that 
\begin{equation}
\frac{1}{\alpha_2(M_{GUT})} - \frac{1}{\alpha_3(M_{GUT})} = \frac{5\kappa}{120}\\.
\end{equation}
It means that playing with the $\kappa$ parameter we can increase 
the GUT scale  $M_{GUT}$. For instance, for $\kappa = 30.2(47.1)$ we obtain  
$M_{GUT} = m_{PL} = 2.4 \cdot 10^{18}~GeV(M_{GUT} = M_{PL}  = 1.2 \cdot 10^{19}~GeV)$.
Nonlocal generalization of nonrenormalizable interaction (13) is 
\begin{equation}
\Delta L_{F\Phi F \Phi2,nl} = \frac{Tr(F_{\mu\nu}\Phi 
V_{\Phi2}(-\Delta_{\mu}\Delta^{\mu})  \Phi F_{\mu\nu})}{\Lambda^2_{\Phi2}} \,.
\end{equation}
For nonlocal Lagrangian (19) as in the case for nonlocal Lagrangian (12) the parameter 
$\kappa$ depends on the scale $\mu$. For the  normalization condition 
$V_{\Phi2}(-M^2_{GUT} ) = 1 $ formulae (15-18) and the numerical estimates are not 
changed.


Let us make our main conclusions. Additional nonlocal interactions (12) or (12,19) allow to 
overcome the standard $SU(5)$ GUT problems with fast proton  decay and wrong 
 Weinberg angle  prediction. Nonlocal generalizations (12,19) 
cure the problems with  ultraviolet behaviour of nonrenormalizable interactions (4,14). 
The price of such modification is the absence of predictive power at least for 
the Weinberg angle $\theta_W$.  The nonlocal scales $\Lambda_{\Phi1}$  
coincide by the order of magnitude with the
GUT scale $M_{GUT}$. In the simplest nonlocal extension of the standard $SU(5)$ GUT 
with $\kappa = 0$ the value of GUT scale is $M_{GUT} \approx 3 \cdot 10^{16}~GeV$. For 
general case with $\kappa \neq 0$ the GUT scale $M_{GUT}$ is an arbitrary. 
It is well known that quantum gravity is nonrenormalizable theory. To 
cure bad ultraviolet properties of quantum gravity 
we have to modify gravity at Planck scale, 
in particular, nonlocal generalization of gravity 
\cite{KRASNIKOV}
leads to superrenormalizable theory \cite{KRASNIKOV,KUZMIN}. Therefore 
the most interesting and natural option is the equality at least by the order of magnitude of 
$M_{GUT}$,  $M_{PL}$ and the nonlocal scale $\Lambda$.

I am indebted to the collaborators of the INR theoretical department 
for discussions and critical comments.

\end{document}